# Formation and applications of nanoparticles in silica optical fibers

*Wilfried Blanc, Bernard Dussardier*

Université Nice Sophia Antipolis, CNRS, LPMC, UMR 7336, Parc Valrose 06108 NICE CEDEX 2, France

corresponding author : bernard.dussardier@unice.fr , tel. : +33 492 076 748

## Abstract

Optical fibers are the basis for applications that have grown considerably in recent years (telecommunications, sensors, fiber lasers, etc). Despite undeniable successes, it is necessary to develop new generations of amplifying optical fibers that will overcome some limitations typical of silica glass. In this sense, the amplifying Transparent Glass Ceramics (TGC), and particularly the fibers based on this technology, open new perspectives that combine the mechanical and chemical properties of a glass host with the *augmented* spectroscopic properties of embedded nanoparticles. This paper is an opportunity to make a state of the art on silica-based optical fibers containing nanoparticles of various types, particularly rare-earth-doped oxide nanoparticles, and on the methods for making such fibers. In the first section of this article, we will review basics on standard optical fibers and on nanoparticle-doped fibers. In the second section we will recall some fabrication methods used for standard optical fibers, and in the third section we will describe how TGC fibers can be obtained. This new generation of fibers, initiated a decade ago, already show very positive outlook and results departing from those obtained previously in homogeneous silica fibers. The next few years should see the emergence of new components based on these optical fibers.

## 1 Introduction : Why optical fibers containing nanoparticles?

In this section we first recall basics on optical fibers, to the benefit of new comers in this field. Then we will motivate the usefulness of inserting nanoparticles inside the core of optical fibers. Finally we discuss the constraints on the nanoparticle dimensions and optical properties for applications. This helps to give directions for the manufacture of TGC fibers.

### 1.1 Basics on optical fibers

Optical fibers are now part of an *everyday technology*, usually associated with telecommunications and high-speed internet. Far from being confined to this single use, optical fibers are deployed in many fields of applications, such as lasers (for marking, machining, cutting, laser remote sensing (LIDAR), etc.) or sensors (civil engineering, gas and oil extraction and storage, temperature senors, etc.).

These applications are based on guided wave optics in optical fibers. In the following section, we will focus more specifically on fibers based on silica ($SiO_2$) working on the principle of total internal reflection (TIR), of which a typical structure is shown in Fig. 1. An optical fiber is a cylindrical optical waveguide made of a *cladding* (refractive index $n_2$) surrounding a central portion called the *core* (refractive index $n_1$). Light is guided by TIR when $n_1 > n_2$. To obtain such an index difference, the fiber core is modified by doping silica with elements such as germanium, phosphorus or aluminium to increase the refractive index. Fluorine or/and boron can be used to decrease the refractive index, either in the core or in the cladding. Both an *index riser* and an *index lowerer* may be used simultaneously in the core or the cladding when specific optical and/or mechanical properties are seeked. The cladding is usually surrounded by a plastic sheath (the *coating*) made of a polymer-based resin to allow for the fiber flexibility and to protect it from mechanical and chemical exposure. Typical diameters of an optical fiber for telecommunication transmission are: plastic coating: 250 µm, optical cladding: 125 µm, core: 6 to 9 µm. The typical refractive index difference is $\Delta n = n_1 - n_2 \sim 3\text{-}5.10^{-3}$. The light of a guided mode is mainly confined to the core, however a substantial part of it may travel in the cladding. Therefore, the characteristics of the whole materials making the waveguide (composition, production conditions, purity, etc.) have a dominant role in the optical properties. The cladding is thick enough so that there is no interaction between the guided mode(s) and the coating. The overall fiber structure is also optimized to lower bending loss, minimize connection/splicing losses. The waveguide characteristics are also designed to meet intermodal and/or chromatic dispersion specifications [1].

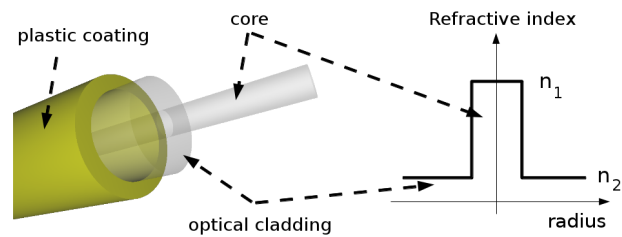

**Fig. 1** Schematic structure of an optical fiber (left) ; schematic refractive index profile (right).

One of the first applications of glass optical fibers was fiberscope. However the poor optical quality of the glass used at that time limited the transmission to a few meters. Charles Kao (Nobel Prize in Physics in 2009) proposed to use silica-based optical fibers as carrier for telecommunications [2]. The subsequent development of manufacturing processes aimed at increasing the transparency by reducing the extrinsic absorption (SiOH, transition metals like iron, etc.) and the intrinsic loss (glass composition and inhomogenities). Within ten years, the attenuation was reduced by four orders of magnitude to reach, in 1979, only 0.2 dB/km (1 % of the transmitted signal remaining after 100 km propagation). By comparison, the attenuation of a lens glass is ~ 2000 dB/km (1 % of the transmitted signal remains after 10 m).

The attenuation curve of a silica fiber is shown in Fig. 2. It has a minimum around 1550 nm. This area of wavelengths is hence favorable for long-distance telecommunications. The absorption peak at 1385 nm is due to the presence of an overtone of vibration modes of O-H bonds. The manufacturing techniques reduce these impurities through chlorine treatment under high temperature to limit the formation of silanol groups in the core. At longer wavelengths, the attenuation increases due to the excitation of vibrational modes of the glass network. This infrared absorption is inherent to the material. It can affect the choice of glass for specific applications : the transparency window is limited to ~ 2 µm in silica, whereas it extends to longer wavelengths in fluoride- or chalcogenide-based glasses. For wavelengths less than 1550 nm, the increase of the attenuation is mainly due to Rayleigh scattering. This depends on the fluctuation of glass density and composition, and is influenced by the manufacture parameters.

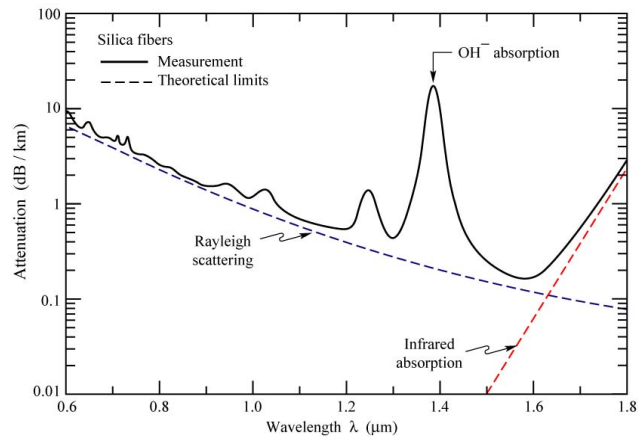

**Fig. 2** Typical attenuation spectrum of silica optical fibers (solid line), and theoretical limits (dashed lines) given by Rayleigh scattering and molecular vibrations (from [3]). Note that current techniques allow for the total reduction of the OH contamination

## 1.2 Nanoparticles for *augmented* fibers

The success of optical fibers based on silica are many: transmission fibers and fiber amplifiers for telecommunications, high-power fiber lasers, sensors, etc. These key applications rely on the qualities of silica glass : mechanical and chemical stability, high optical damage threshold, low cost, etc. However, silica glass has certain characteristics which may make it less efficient compared to other types of glass, particularly in some potential applications using rare earths or transition metals luminescent ions. Indeed, the luminescence properties of rare earth and transition metal ions are linked to their electronic structure. The optically active electrons in transition metals belong to their *d* layer, wheras the *f* layer electrons of rare-earth elements offer most of the electronic transitions of interest for optical fiber applications (like the $^4I_{13/2} \rightarrow {}^4I_{15/2}$ emission transition of $Er^{3+}$ at 1.5 µm or the $^2F_{5/2} \rightarrow {}^2F_{7/2}$ transition for $Yb^{3+}$ main emission around 1 µm). The spectroscopic properties of these luminescent ions (oscillator strengths, energy of the electronic levels, electron-phonon coupling, etc.) strongly depend on the arrangement and the nature of the surrounding atoms. Silica has a relatively high phonon energy that can result in non-radiative de-excitations among energy levels of the luminescent ions : this may affect the emission quantum efficiency of certain optical transitions [4].Furthermore, the low solubility of rare-earth ions in silica glass leads to the formation of aggregates, or clusters, that leads to luminescence quenching phenomena at concentrations greater than 100 ppm [5].

Alternative glasses have also been proposed as improved hosts for rare-earth elements [6]. These glasses would provide better quantum efficiency or an improvement emission bandwidth to some optical transitions of particular interest. The icon example was the $Tm^{3+}$-doped fibre amplifier (TDFA) for telecommunications in the S-band (1.48-1.53 µm) [7], for which low phonon energy glasses have been developed: fluorides [8], chalcogenides [9], and even oxides [10,11]. Despite the greatest efforts toward that aim, very few, if any, implementations of these active alternative glasses have been proposed in applications. Indeed, these glasses suffer bad compatibility with most commercial applications. The main causes are : high fabrication cost, low reliability, difficult connection to silica components and, in the case of fibre lasers, low optical damage threshold and bad resistance to heat. Silica glass is the only material able to stand the operation at high optical power in fibre devices. Therefore the choice of vitreous silica for the active fibre material, or for the host material (see below) is of critical importance.

In the case of transition metal ions, the symmetry of the occupied site is critical for the emission properties. For example, the de-excitation between two electronic levels of $Ni^{2+}$ is radiative if ions are in an octahedral environment while it is non-radiative in other sites with different symmetry in the glass [12]. The silica glassy matrix offers sites having a structure or symmetry that is detrimental to the fluorescence of transition metal ions, unlike sites that are present in some crystals. Also, the great number of different sites provided by an amorphous host such as glass results in the inhomogeneous enlargement of the emission band: this causes a decrease of the emission cross-section, detrimental for some applications such as fiber lasers.

To overcome these limitations, it was proposed to develop silica-based optical fibers containing nanoparticles whose composition shall be chosen to benefit the intended application. It is well known that the spectroscopic properties of luminescent ions are sensitive to the characteristics of their so-called *local environment*, that is to say the composition and the structure (symmetry) of the few first coordination shells around it [13,14] (see section 3). Such fibers with *augmented* properties would combine the advantages of silica (transparency, cost, chemical and mechanical durability, etc.) and the specific properties provided by the nanoparticles. For example, the formation of crystalline phases was studied in a fiber with multicomponent starting composition in the aim of increasing non-linear effects [15,16].

The incorporation of the metal nanoparticles (MNP) in glass fibers is also of interest. For example, fibers obtained from an oxyfluoride starting glass containing silver nanoparticles have been proposed to develop a new light source at 650 nm [17]. The confinement of free electrons at the metal / dielectric interface may generate, under the effect of a light wave, a resonant excitation of conduction electrons of the metal giving rise to localized plasmonic modes. This may induce a dramatic increase in linear optical responses and non-linear materials located in the immediate vicinity of these structures. Enhanced luminescence from rare earths ions has been reported in a silica film due to the presence of nanoparticles of gold or silver [18-20]. However, recent modeling results demonstrate that the enhancement of the fluorescence is strongly dependent on the distance between rare earth ions and metal particles [20]. An other major difficulty to associate MNP to optical fibers is related to the manufacturing and shaping temperatures (> 2000 ° C), which pose the problem of conservation of nanostructures in the fiber. Yet recent studies describe the manufacture of optical fibers doped with nanoparticles of gold or silver, to exploit their nonlinear optical properties and photoluminescence [21-24].

The following section deals more specifically with luminescent ions embedded in dielectric nanoparticles to taylor their spectroscopic properties. Although nanoparticles may be either amorphous or crystalline, we will use in the following section the generic term *transparent glass ceramic* (TGC) fiber to designate fibers containing dielectric nanoparticles.

## 1.3 Constraints on the nanoparticle optical characteristics

The presence of nanoparticles in the core of TGC optical fibers leads to optical loss due to the Rayleigh scattering. This must be minimized for pratical applications. To determine the acceptable size of these nanoparticles for a given application, we use the following formula to estimate Rayleigh loss [25] :

$$\alpha_{Rayleigh}(dB/m) = 4.34 \times C_{Rayleigh} \times N \times \Gamma \tag{Eq. 1}$$

where $N$ is the particle density (m$^{-3}$), $\Gamma$ is the overlap factor between the guided mode and the region of the fiber containing the nanoparticles and $C_{Rayleigh}$ (m$^2$) is the Rayleigh scattering coefficient such as [25]:

$$C_{Rayleigh} = \frac{(2\pi)^5}{48} \times \frac{d^6}{\lambda^4} \times n_m^4 \times \left(\frac{n_n^2 - n_m^2}{n_n^2 + 2n_m^2}\right)^2 \tag{Eq. 2}$$

where $d$ is the diameter of the nanoparticle, $n_n$ and $n_m$ are the refractive indices of the matrix and of the nanoparticle at the wavelength under consideration, respectively. Fig. 3 represents the Rayleigh loss at 1550 nm, corresponding to the particle density $N=10^{20}$ m$^{-3}$, and the overlap factor $\Gamma=1$, versus the particle size, for four values of the index difference $\Delta n = n_m - n_n$. For comparison, $\Delta n = 2$ corresponds to the index difference between silica and silicon particles. The TGC fibers are considered for applications requiring fiber lengths typically few meters to tens of meters, unlike line fibers that must transmit over hundreds of kilometers. Therefore a threshold of 0.1 dB/m loss is considered acceptable. It follows that the particle size should be less than 100 nm, and typically ~10 nm, with the index difference $n_m - n_n$ as small as possible (Fig. 3). This result is in agreement with the characteristics of the first bulk TGC studied for an optical application: a silicate matrix containing Pb$_x$Cd$_{1-x}$F$_2$ particles of average size 20 nm and an index difference $\Delta n = 0.3$ [26,27]. Later, four criteria were established to meet the transparency requirements for propagation on short lengths (<1 km): i) the particle size must be less than ~15 nm, ii) the distance between particles must be comparable to the particle size, iii) the size distribution must be narrow, and iv) the particles must not cluster together [28].

Furthermore, in the particular case of luminescent ions incorporated within dielectric nanoparticles, many reports show that their spectroscopic properties can be greatly enhanced if the particle diameter is ≤ 20 nm [29]. Thus, the transparency requirements can be met while maintaining the control of the nanoscale environment around luminescent ions within the nanoparticles. Note that the index difference between silica and most oxide nanoparticles may be less than 0.3: this would loosen the constraint on the particle size.

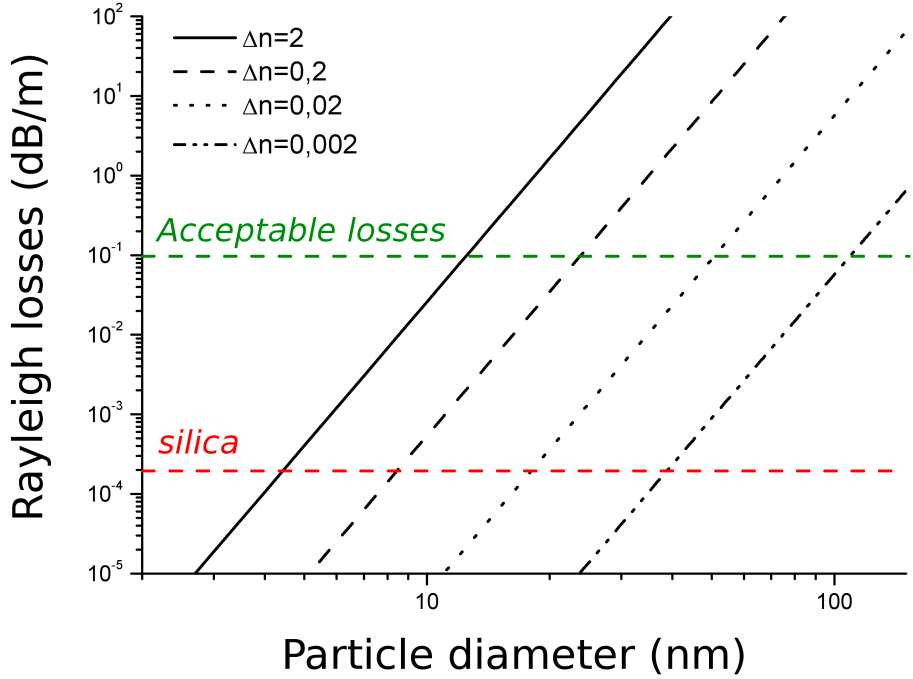

**Fig. 3** Rayleigh losses, calculated by (Eq. 1), associated with the presence of nanoparticles ($\lambda$ = 1550 nm, $N=10^{20}$ particles/m$^3$, $\Gamma=1$) in function of the nanoparticle diameter, for different value of the index contrast $\Delta n = n_m - n_n$ (see text). Red and green dashed lines correspond to the minimum attenuation in the silica and acceptable attenuation threshold, respectively, as defined in the text.

## 2  Some usual methods for making optical fibers

TGC optical fibers are prepared by different processes that were derived from standard methods originally developed for making *homogeneous* fibers, that is to say not containing nanoparticles. These techniques are synthetically presented in this section. They share the same two main steps: preparing a glass containing the nanoparticles or the elements necessary to form the nanoparticles, and then drawing it into a fiber. The preparation of TGC fibers is discussed in the next section. The optical fibers obtained by the Rod-In-Tube, double crucible, or MCVD (Modified Chemical Vapor Deposition) have a core-cladding structure as shown in Fig. 1. Fibers can also be obtained by stretching a single glass into the shape of a cane, i.e. with no cladding. For example, these *bulk* fibers are produced to evaluate the potential of a glass or TGC to be drawn into a fiber.

### 2.1  The *Rod-in-Tube* process

The Rod-in-Tube (RIT) method, shown schematically in Fig. 4(a), was historically the first method used to make optical fibers [30]. A glass rod, the future core, is inserted into a glass tube, the future cladding. The composition of these two elements should be selected according to two criteria: (i) cladding refractive index $n_2$ lower than the core refractive index $n_1$, and (ii) compatible thermo-mechanical properties with the drawing process. In particular, their coefficients of expansion must be very close. This assembly is drawn into an optical fiber under heating and stretching. This process allows *a priori* to convert any glass into the core of an optical fiber, under the condition of finding a suitable complementary glass to form the cladding. Despite its apparent simplicity, this method is seldom used because the obtained fibers are lossy (as compared to other methods) due to residual surface defects and contamination on the cladding tube and the core rod.

## 2.2 The *Double-Crucible* method

This method is similar to the previous one as it uses two starting materials. However, in the case of the double crucible method, the fiber is obtained from the flow of two molten glasses through a specially designed nozzle. The core of the glass is melted in the inner crucible, that of the cladding in the outer crucible (Fig. 4(b)). This method provides access to a wide range of materials, but has a high risk of contamination by impurities from the crucible at the core-cladding interface.

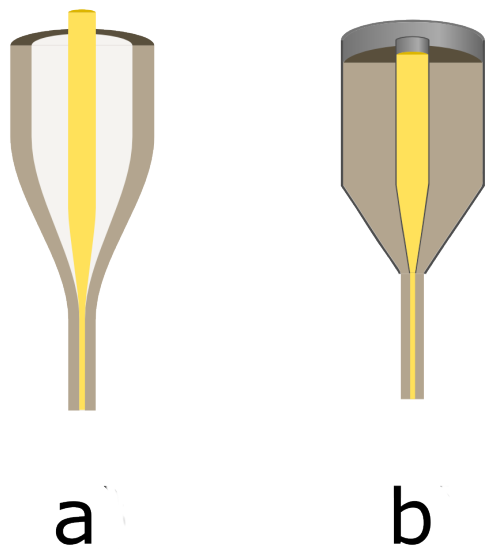

**Fig. 4** Principle (a) Rod-in-Tube and (b) Double-Crucible processes

## 2.3 MCVD method, solution doping and fiber drawing

Among the various methods based on the Chemical Vapor Deposition (CVD), the Modified Chemical Vapor Deposition (MCVD) is the most flexible technique. It is mostly used for the production of specialty optical fibers based on silica. MCVD was developed in the 1970s at AT&T Bell Labs [31], and further developed within few years [32]. The production steps are shown schematically in Fig. 5. This process is based on the successive deposition of vitreous layers inside a silica glass tube. The composition of the layers is determined by those of the reactive gas (mainly $SiCl_4$, $GeCl_4$ and $POCl_3$ and a fluorine carrier like $C_2F_6$, $SF_6$ or $SiF_4$) carried in the tube by oxygen. Additive elements, among which rare-earth elements, aluminium and other modifiers, are provided during the operation of solution doping [33]. The solution doping consists in soaking a porous silica layer (formed by MCVD at relatively low temperature) with an alcoholic or aqueous solution of salts of metal chlorides ($ErCl_3$, $AlCl_3$, etc.). After removing of the solution and drying, the porous layer is densified and vitrified at high temperature (up to 1800 °C). The tube is closed during the so called *collapse* step (at very high temperature > 2000 °C) and forms a rod called *preform* whose diameter is ~1 cm and its length is ~ 50 cm (dimensions usually obtained in a laboratory). This preform is then drawn into an optical fiber of diameter 125 μm. The stretching is obtained by introducing the preform in a furnace to 2000 °C in order to exceed the softening temperature of silica. Note that alternative doping using gas phases of rare-earth compounds have been proposed [34].

This method allows to obtain materials of very high purity (related to the purity of the precursors). The total amount of modifiers of pure silica is limited to a few mole percent to ensure the compatibility of core and cladding glasses, and because devitrification occurs and loss increases at higher concentrations. Unlike the methods described above, here the layer deposition passes, the densification of the soaked porous core layer, the collapse and the drawing steps require many thermal cycles to temperatures up to more than 2000 °C. A specific point of a fiber has been submitted to these thermal cycles for only few seconds to few minutes at each pass or step. This remark is important for the preparation of phase separated TGC fibers (next section).

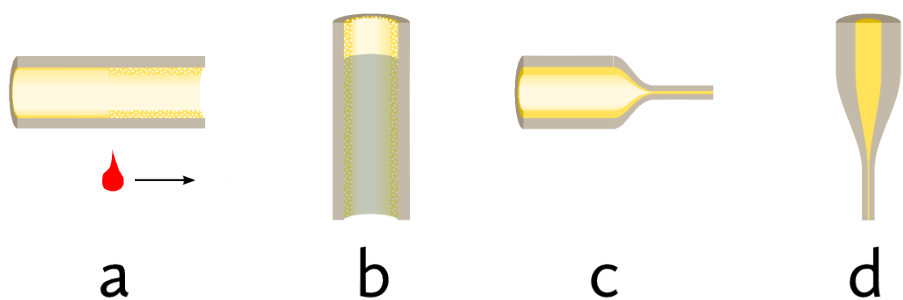

**Fig. 5** Principle of the preparation of the rare-earth-doped fibre by MCVD and solution doping: (a) deposition of a porous silica layer within a rotating silica tube, (b) solution doping, (c) collapsing and (d) fiber drawing. The glass layer (a) is obtained by the oxidation reaction of $SiCl_4$ and $O_2$ into $SiO_2$ and $Cl_2$. The heat source is usually the flame of a translating oxygen-hydrogen torch (represented by the flame in (a)), that allows for deposition along the whole substrate tube length.

## 3    Transparent glass ceramic optical fibers

The first report on the preparation of a vitroceramic with optical properties was published in 1975 [35]. The particles were lead fluoride ($PbF_2$) microcrystals in an oxide matrix ($SiO_2$, $GeO_2$, $P_2O_5$, $B_2O_3$, $TeO_2$). The particle mean size was ~10 μm: obviously the vitroceramic had high transmission losses. Since the early 2000s, several compositions and several manufacturing processes were studied to develop fibers containing particles. The methods described in the preceeding section are used to prepare preforms and to draw them into optical fibers. In most cases, fibers do not contain nanoparticles yet. In this section we present the formation of nanoparticles in fibers which can be formed in the preform before it is drawn into a fiber, or on the application of an additional thermal treatment on the fiber after it is drawn, called *post heat-treatment*.

Until recently, this field was emergent, and most efforts were focused on solving the many issues caused by the composition and the structure in the bulk, to the nanometric scale. This is why the demonstration of applications based on TGC optical fibers remained marginal up to very recently (see next sub-section). In addition, the development of knowledge in this area suffers from a lack of precise, reliable and quantitative characterization techniques applicable on the TGC in the core of optical fibers (the core volume is ~1% of the fiber volume). And the TGC itself contains 10-to-100-nm nanoparticles occupying typically only ~1 to 10% of the fiber core. Because of the difficulty to manipulate and prepare such small sample, most reports concern only bulk preform samples. Often these reports ignore the effects of the fiber drawing stage on the final properties of the nanoparticles, as it was reported elsewhere, recently [36]. Here we restrict our discussion to the results obtained in the aim of developping TGC core optical fibers with specific properties for identified applications. The sought interesting properties comprise, but are not limited to, broadband gain, high quantum efficiency and resistance to photodarkening.

## 3.1 Ceramization of optical fibers

Like for bulk glass ceramics, post heat-treatment of the optical fiber was used to form nanoparticles in oxy-fluoride or oxide fibers. In the case of oxyfluoride fibers, a $Nd^{3+}$-doped glass was prepared by the double crucible method [37,38]. By post heat-treatment of the fiber at 460 °C for 5 h, crystals of composition 60 $PbF_2$ – 20 $YF_3$ – 20 $CdF_2$ (size less than 10 nm) were formed, containing all the $Nd^{3+}$ ions. In another study, Er-doped and Yb-Er-codoped fibers have been drawn (at 720-740 °C) from oxyfluoride glass compositions, with less then 15 mol% of fluorides [39]. After heating the fiber at 700 °C for 50 h, fluoride and oxyfluoride crystalline nanoparticles having a diameter of 25 ± 5 nm were identified. Note that the rare-earth ions were encapsuled only in oxyfluoride crystallites.

In the case of *oxide-only* glasses, the thermal post-treatment was applied to nickel- or chromium-doped almunino-gallo-silicate fibers prepared by the *Rod-In-Tube* method [40,41]. $Ni^{2+}$-doped alumino-gallate crystals are obtained after heating for 2 h at 850 °C. In the case of a nickel- and antimony-codoped gallo-germano-silicate core rod, the formation of nanoparticles is obtained by applying a heat treatment for 1.5 h at 840 °C [42]. According to studies on the bulk material, phases are precipitated as $LiGa_5O_8$ and γ-$Ga_2O_3$, with an average size of about 6 ± 2 nm. One note that in this case, crystallites preferably formed at the interface between the core and the silica cladding: it shows that the post-heat treatment approach does not always provide a beneficial transformation to the bulk of the core volume.

Several amplifying cerammed TGC-core fibers were implemented on laboratory devices such as fiber lasers [37] or fiber amplifiers [40].

## 3.2 Phase-separated TGC by MCVD

It has also been proposed to synthesize nanoparticles *in-situ* directly into the preform using the MCVD process and the solution doping method, or methods derived from the latter. Here no post-heat treatment on the fiber would be necessary, in principle. Instead the approach takes advantage of the many temperature-cycles and/or the bringing of externaly prepared nanoparticles. The absence of post-heat treatment would be a critical advantage of this approach for an industrial development.

One of the most crucial steps is the formulation of the doping solution (or alternatively, the particle liquid suspension). The solution/liquid suspension may contain (i) nanoparticles already pre-formed by chemistry (such as $Er^{3+}$-doped $Al_2O_3$ or $SiO_2$ nanoparticles [43]), (ii) glass particles obtained from grounding of a bulk glass sample ($Co^{2+}$-doped $ZnAl_2O_4$ crystals, size 200-300 nm [44]) or (iii) chemical elements that will trigger the *phase-separation* during the thermal cycles that are inherent of the MCVD process. The first route leads to rather homogeneous glass, i.e. the particles have totally or partially dissolved in the matrix, although some benefit of this approach was observed (radiation resistance [43]). The second was reported once, to the best of our knowledge.

This third route was systematically explored by us: porous core layers (either $SiO_2$, $SiO_2$-$GeO_2$ or $SiO_2$-$P_2O_5$-$GeO_2$ compositions) using MCVD is soaked with an alcoholic or aqueous solution containing erbium and alkaline-earth salts. The studied alkaline-earth ions are magnesium ($Mg^{2+}$), calcium ($Ca^{2+}$) and strontium ($Sr^{2+}$) ions. We obtained 50-to-100-nm nanoparticles that were enriched with erbium and magnesium [45]. Note that no additional pre- or post-heat treatment was necessary, the TGC is directly available after the fiber drawing stage. Other groups observed the growth of 5-to-8-nm particles enriched in yttrium, ytterbium and aluminum synthesized via this third way. However these nanoparticles are observed in the fiber only if the preform is submitted to a *pre-heat treatment* (3 hours at 1450 °C) before the fiber drawing [46]. Further, particles that were observed as crystalline in the preform became amorphous during the drawing stage.

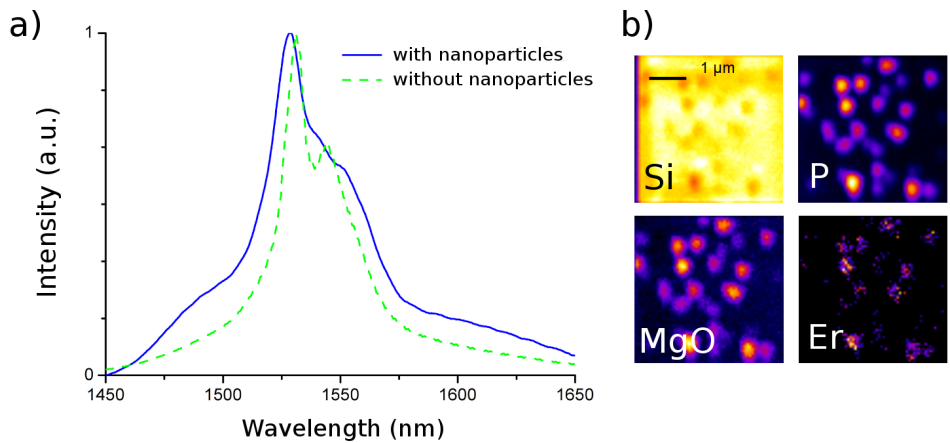

**Fig. 6**(a) Emission spectra of $Er^{3+}$ measured in an optical fiber without Mg (green line, no nanoparticles in the fiber) or with Mg (blue line, presence of nanoparticles in the fiber), the excitation wavelength is 980 nm. (b) Spatial distributions of Si, P, MgO and Er elements measured by mass spectrometry (for details, see [48]).

The partition of the luminescent ions in nanoparticles (that is to say, the specific location of luminescent ions inside the nanoparticles, and not in the matrix) results in a change of spectroscopic properties compared to those in the parent glass. For example, Fig. 6(a) shows the broadening of the emission spectrum of $Er^{3+}$ obtained from fiber containing nanoparticles induced by the presence of magnesium [47]. It is compared with the emission from a fiber without magnesium. Fig. 6(b) shows the distribution of Si, MgO, P and Er elements measured by mass spectrometry. These three elements are located in the nanoparticles only [48]. This result is very important: it shows that this technique may open the way to reproducible spectroscopic-engineering in rare-earth-doped optical fibers. Indeed, when erbium ions are distributed among several discret local environments (so-called "sites") [49], or over a continuous distribution of sites, then the gain curve is characterized by a complex inhomogeneous broadening. The gain curve is expected to be noticeably different compared to that in standard erbium doped fibers (EDF) based on alumino-silicate glass. With this type of EDF, broad, and relatively flat gain curves may be achieved within a same fiber. In the concept shown here, erbium ions are subjected to a variety of compositon and local structure. Although it is difficult to estimate how much broadening is achievable, the gain curve shape is expected to show less ripple amplitude.

This would give a lesser gain excursion accros the band of interest. In particular, in the C-band of the telecommunications the typical "ripples" of erbium gain in alumino-silicate fibers (peaks at 1.53 and 1.55 µm, dip at 1.54 µm, and down slope beyond 1.56 µm) would be smeared out or minimized. This would benefect to the lowering of EDFA cost, by allowing for the simplification of the output gain flattening filter (GGF).

# 4  Conclusion

To overcome some limitations of silica glass (gain bandwidth, luminescence efficiency, non-linearity, ...), it is proposed to develop a new generation of optical fibers based on core containing nanoparticles. In this paper, we focus on recent progress on luminescent ions doped oxide nanoparticles of interest for amplifying optical fibers. This technology opens new perspectives that combine the mechanical and chemical properties of a glass host to the *augmented* spectroscopic properties of nanoparticles. Given the emergence of this new type of fiber, most studies of such materials aim at demonstrating the potential of a glass and a method for manufacturing an optical fiber with interesting properties. Two main routes are investigated to obtain nanoparticles in fibers : (i) directly during the manufacturing process of the fiber, (ii) afterwards by applying an additional heat treatment to the fiber. Up to now, the demonstration of components based on such optical fibers remains marginal. However, the next few years should see the emergence of new components based on these optical fibers.